\documentclass[aps,showpacs]{revtex4}
\usepackage{graphics,epsfig,psfig}

\begin{document}

\def\ds#1{\frac{\partial^2 #1}{\partial\omega'^2}|_{\omega'=0}}
\def\calm{{\mathcal M}}
\def\dcalm{\Delta{\mathcal M}}

\title{New predictions for generalized spin polarizabilities from
heavy baryon chiral perturbation theory}
\author{Chung-Wen Kao$^1$, Barbara Pasquini$^2$, and Marc Vanderhaeghen$^3$}

\affiliation{$^1$Department of Physics, North Carolina State University, 
Raleigh, NC27695-8202,USA}
\affiliation{$^2$ Dipartimento di Fisica Nucleare e Teorica, Universit\`a degli
Studi di Pavia; \\ 
INFN, Sezione di Pavia, Pavia, Italy, and ECT*, Villazzano (Trento), Italy}
\affiliation{$^3$Thomas Jefferson National Accelerator Facility, Newport News, VA 23606,USA and\\
Department of Physics, College of William and Mary, Williamsburg,VA 23187,USA}

\date{\today}

\begin{abstract}
We extract the next-to-next-to-leading order results for  
spin-flip generalized polarizabilities (GPs) of the nucleon from
the spin-dependent amplitudes 
for virtual Compton scattering (VCS) at ${\cal O}(p^4)$
in heavy baryon chiral perturbation theory. 
At this order, no unknown low energy constants enter the theory, 
allowing us to make absolute predictions for all spin-flip GPs. 
Furthermore, by using constraint equations between the GPs due to 
nucleon crossing combined with charge conjugation symmetry 
of the VCS amplitudes, we get a 
next-to-next-to-next-to-leading order prediction for one of the GPs. 
We provide estimates for forthcoming double polarization experiments 
which allow to access these spin-flip GPs of the nucleon. 
\end{abstract}
\maketitle

\section{Introduction}
\label{sec1}

Over the past decade, the virtual Compton scattering (VCS) process on
the nucleon, accessed through the $e p \to e p \gamma$ reaction, has
become a powerful new tool to study the internal structure of the
nucleon both at low and high energies (see \cite{GV98} for a review). 
At low virtuality and energy, the outgoing real photon plays the role
of an applied quasi-static electromagnetic field, and the VCS process
measures the response of the nucleon to this applied field, which can
be parametrized in nucleon structure quantities, termed 
generalized polarizabilities (GPs) \cite{GLT}. 
In this case, the virtuality of the initial photon can be dialed so as
to map out the spatial distribution of the electric polarization, the
magnetization, or the spin densities of the nucleon~\cite{spatial}.
Because the chiral dynamics plays a dominant role in this regime, 
at low virtuality,  
Heavy Baryon Chiral Perturbation Theory (HBChPT) provides a natural
way to study the GPs of the nucleon.
\newline
\indent
The spin-dependent VCS amplitudes in HBChPT at order
${\cal O}(p^3)$ have been calculated in Refs.~\cite{HHKS,HHKD}. 
The HBChPT calculation for the spin-dependent VCS amplitudes has 
recently been extended to ${\cal O}(p^4)$ in Ref.~\cite{KV02}. 
At this order, no unknown low-energy constants enter
the calculations so that one can extract 
the next-to-leading order (NLO) results for the 
spin-flip GPs of the nucleon, without free parameters. 
In Ref.~\cite{KV02}, it was possible to extract the next order corrections to 
three of the seven spin-flip 
GPs in HBChPT by identifying the terms linear in the final photon energy 
in a low-energy expansion (LEX) of the spin-dependent VCS amplitudes. 
\newline
\indent
In this paper, we demonstrate that by performing higher order
expansions of the VCS amplitude with respect to the final photon energy, 
i.e. by including the terms quadratic in the final photon energy, 
one can furthermore 
extract the NNLO results of three more spin-flip GPs and 
even make a $N^{3}LO$ prediction for one particular GP. 
In this way we are able to obtain the next order corrections to 
all seven spin-flip GPs in HBChPT, without free parameters.   
\newline
\indent
On the experimental side, the measurement of the VCS process
became only possible in recent years, with the advent of high
precision electron accelerator facilities, as these experiments
involve precision measurements of small cross sections. At low photon
energy and virtuality, 
first unpolarized VCS observables have been measured  
at the MAMI accelerator \cite{Roc00} at a virtuality $Q^2$ = 0.33 GeV$^2$, 
and recently at JLab \cite{Lav04} at higher virtualities, 
$1 < Q^2 < 2~$GeV$^2$, and data are under analysis at MIT-Bates \cite{Mis97}.
All those experiments measure two combinations of GPs. 
To disentangle the four independent spin-flip GPs, requires to perform 
in addition double polarization experiments. 
In this paper, we show predictions for 
such double polarization asymmetries and 
demonstrate their sensitivity to the spin-flip GPs.  
\newline
\indent
The outline of this paper is as follows. We start in 
Section~\ref{sec2} by reviewing how the virtual Compton scattering 
amplitude at low photon energies can be expanded in terms of GPs. 
We subsequently review the status of the 
HBChPT calculations for the spin-flip GPs. 
In Section~\ref{sec3}, we outline how we can extract four spin-flip GPs at 
next order in HBChPT by performing a quadratic expansion in the final 
photon energy. In Section~\ref{sec4}, we compare our 
predictions for the four independent spin-flip GPs at next order in HBChPT 
with the leading order (non-zero) expressions. 
Furthermore, we demonstrate the sensitivity 
of double polarization asymmetries to the spin-flip GPs. 
Finally we give our conclusions in Section~\ref{sec5}. Technical details 
on crossing symmetry relations for the VCS amplitudes can 
be found in an Appendix.

\section{Virtual Compton Scattering and Generalized Polarizabilities}
\label{sec2}

We begin by specifying our notation for the VCS process:
\begin{equation}
\gamma^{*}(\varepsilon_{1},q)+N(p_{1})\rightarrow\gamma(\varepsilon_{2},q')+N(p_{2}).
\end{equation} 
The VCS amplitudes are calculated in the {\it c.m.} frame and 
we choose the Coulomb gauge \cite{KV02}.
In the VCS process, the initial spacelike photon is characterized 
by its four momentum $q=(\omega,\vec{q})$, virtuality $Q^2 \equiv - q^2$, 
and polarization vector $\varepsilon_{1}=(0,\vec{\varepsilon}_{1})$.
The outgoing real photon has four momentum 
$q'=(\omega'=|\vec{q'}|,\vec{q'})$ and 
polarization vector $\varepsilon_{2}=(0,\vec{\varepsilon}_{2})$.
We define $\bar{q}\equiv|\vec{q}|$, and denote $\theta$ as the
scattering angle between virtual and real photons, i.e., 
$\cos\theta=\hat{q}\cdot\hat{q}^{\, '}$.
The polarization vector $\vec{\varepsilon}_{1}$ of the virtual photon 
can be decomposed into a longitudinal component 
$\vec{\varepsilon}_{1L}=(\vec{\varepsilon}_{1}\cdot\hat{q})\hat{q}$ and 
a transverse component $\vec{\varepsilon}_{1T}$. 
For further use, we introduce the virtual photon energy in the limit 
$\omega' = 0$~:
\begin{eqnarray}
\omega_{0}\equiv \omega(\omega'=0,\bar{q})&=&
M_{N}-\sqrt{M_{N}^{2}+\bar{q}^{2}},
\nonumber\\
&=& -{\bar{q}^2} / (2 M_{N}) + {\mathcal O}(1/M_{N}^{3}),
\label{eq:omegao}
\end{eqnarray}
with $M_N$ the nucleon mass, and where the 
last line in Eq.~(\ref{eq:omegao}) indicates the heavy baryon expansion.
\newline
\indent
The VCS amplitude ${\cal M}_{VCS}$ 
can be expressed in terms of twelve structure functions as \cite{HHKS}~: 
\begin{eqnarray}
{\cal M}_{VCS} \,&=&\, i e^2 \chi_f^{\dagger} \left\{
(\vec{\varepsilon}_{2}^{\, *}\cdot \vec{\varepsilon}_{1T})A_{1}
+(\vec{\varepsilon}_{2}^{\, *}\cdot\hat{q})
(\vec{\varepsilon}_{1T}\cdot\hat{q'})A_{2} 
+\, i\vec{\sigma}\cdot(\vec{\varepsilon}_{2}^{\, *}
\times\vec{\varepsilon}_{1T})A_{3}
+i\vec{\sigma}\cdot(\hat{q'}\times\hat{q})
(\vec{\varepsilon}_{2}^{\, *}\cdot \vec{\varepsilon}_{1T})A_{4} 
\right. \nonumber \\
&+&\, i\vec{\sigma}\cdot(\vec{\varepsilon}_{2}^{\, *}\times\hat{q})(\vec{\varepsilon}_{1T}\cdot\hat{q'})A_{5}
+i\vec{\sigma}\cdot(\vec{\varepsilon}_{2}^{\,*}
\times\hat{q'})(\vec{\varepsilon}_{1T}\cdot\hat{q'})A_{6} 
-\, i\vec{\sigma}\cdot(\vec{\varepsilon}_{1T}\times\hat{q'})
(\vec{\varepsilon}_{2}^{\,*}\cdot\hat{q})A_{7} 
-i\vec{\sigma}\cdot(\vec{\varepsilon}_{1T}\times\hat{q})
(\vec{\varepsilon}_{2}^{\,*}\cdot\hat{q})A_{8} \nonumber \\
&+&\left. \, (\vec{\varepsilon}_{1L}\cdot \hat{q}) \left[
(\vec{\varepsilon}_{2}^{\,*}\cdot\hat{q}) A_{9}
+i\vec{\sigma}\cdot(\hat{q'}\times\hat{q})
(\vec{\varepsilon}_{2}^{\,*}\cdot\hat{q}) A_{10}  
+ \, i\vec{\sigma}\cdot(\vec{\varepsilon}_{2}^{\,*}\times\hat{q}) A_{11}
+i\vec{\sigma}\cdot(\vec{\varepsilon}_{2}^{\,*}\times\hat{q'}) A_{12}
\right] \right\} \chi_i \, . 
\label{eq:vcsampl}
\end{eqnarray}
To extract the GPs, we first calculate the complete fourth order VCS 
amplitudes $A_i$ in HBChPT, and subsequently separate the amplitudes
in a Born part $A_{i}^{Born}$, and a non-Born part $\bar{A}_{i}$ as~:
\begin{eqnarray}
A_{i}(\omega',\bar{q},\theta)=A_{i}^{Born}(\omega',\bar{q},\theta)
+\bar{A}_{i}(\omega',\bar{q},\theta),
\end{eqnarray}
In the Born process, the virtual photon is absorbed on
a nucleon and the intermediate state remains a nucleon. 
We calculate them, following the definition of Ref.~\cite{GLT}, 
from direct and crossed Born diagrams with the
electromagnetic vertex given by~:
\begin{equation}
\Gamma^{\mu}(q^2)=F_1(q^2)\gamma^{\mu}
+F_2(q^2)i\sigma^{\mu\nu}\frac{q_{\nu}}{2M_{N}},
\end{equation}
where $F_1 (F_2)$ are the nucleon Dirac (Pauli) form factors respectively.
The non-Born terms contain the information 
on the internal structure of the nucleon. 
In particular, we are interested in this work 
in the response of the nucleon to an 
applied quasi-static electromagnetic dipole field. 
This is accessed by performing 
a low energy expansion (LEX), in the outgoing photon energy,  
of the non-Born VCS amplitudes, and by selecting the  
term which is linear in the outgoing photon energy, given by~:
\begin{equation}
\left[\frac{\partial \bar{A}_{i}}{\partial \omega'}(\omega',\bar{q},\theta)\right]_{\omega'=0}
={\cal S}_{i}(\bar{q})+{\cal P}_{i}(\bar{q})\cos\theta .
\label{eq:linterm}
\end{equation}
In Eq.~(\ref{eq:linterm}), 
the nucleon structure quantities ${\cal S}_{i}$ and 
${\cal P}_{i}$ can be expressed in terms of the GPs of the
nucleon, as introduced in~\cite{GLT}, 
which are functions of $\bar q$ and which are denoted by
$P^{(\rho' \, L', \rho \,L)S}(\bar q)$.
In this notation, $\rho$ ($\rho'$) refers to the
electric ($E$), magnetic ($M$) or longitudinal ($L$) nature of the initial
(final) photon, $L$ ($L'$) represents the angular momentum of the
initial (final) photon, and $S$ differentiates between the
spin-flip ($S=1$) and non spin-flip ($S=0$)
character of the transition at the nucleon side.
Restricting oneself to a dipole transition for the final photon
(i.e. $L'$ = 1), angular momentum and parity conservation leads to
3 scalar and 7 spin GPs \cite{GLT}.
The 7 spin GPs are obtained as \cite{Mainz1}~:
\begin{eqnarray}
P^{(M1,L2)1}&=&
\frac{-2\sqrt{2}}{3\sqrt{3}} \frac{1}{\bar{q} \, \omega_0}
\sqrt{{{M_N} \over E_N}} 
{\cal S}_{10},\,\,\,
P^{(L1,L1)1} = \frac{-2}{3} \frac{1}{\omega_0} \sqrt{{{M_N} \over E_N}} 
{\cal S}_{11},\,\,\,\, 
P^{(M1,L0)1} = \frac{2}{\sqrt{3}} \frac{\bar{q}}{\omega_0} 
\sqrt{{{M_N} \over E_N}} 
\left[{\cal S}_{12} -\frac{2}{3} {\cal S}_{10}\right], 
\nonumber \\
P^{(L1,M2)1}&=&-\frac{\sqrt{2}}{3 \bar{q}^2} \sqrt{{{M_N} \over E_N}} 
{\cal S}_{8},\,\,\,
P^{(M1,M1)1}=\frac{2}{3 \bar{q}} \sqrt{{{M_N} \over E_N}} 
\left[{\cal S}_{7} -{\cal S}_{4}\right],\nonumber \\
\hat{P}^{(M1,2)1}&=&-\frac{4}{3\sqrt{10} \bar{q}^3} \sqrt{{{M_N} \over E_N}} 
\left[{\cal S}_{4}+{\cal S}_{7}-{\cal S}_{10}\right],\,\,\,\,
\hat{P}^{(L1,1)1}=
-\frac{2\sqrt{2}}{3\sqrt{3} \bar{q}^2} \sqrt{{{M_N} \over E_N}} 
\left[{\cal S}_{3}+\frac{1}{2}{\cal S}_{8}-{\cal S}_{11}\right] ,
\label{gp}
\end{eqnarray}
where $E_N = \sqrt{M_{N}^{2}+\bar{q}^{2}}$, and 
where the GPs denoted by $\hat P$ correspond with 
mixed electric and longitudinal
multipoles, as introduced in~\cite{GLT}.
It has been shown \cite{Mainz1} that
nucleon crossing combined with charge conjugation symmetry of the VCS
amplitudes provides 3 constraints among the 7 spin GPs~:
\begin{eqnarray}
{\cal S}_{4}=0, 
\hspace{0.75cm}
{\cal S}_{3}=\frac{\bar{q}}{\omega_0} {\cal S}_{7}, 
\hspace{0.75cm}
{\cal S}_{10}-{\cal S}_{12}=\frac{\bar{q}}{\omega_0} {\cal S}_{11},
\label{con2}
\end{eqnarray}
leaving 4 independent spin GPs. 
By expanding $\omega_{0}$ as in~(\ref{eq:omegao}), 
it is obvious that the relations (\ref{con2}) 
connect quantities of different order in the heavy baryon
expansion. These constraints have been verified in HBChPT
in Ref.~\cite{KV02}. 
\newline
\indent
By calculating the VCS amplitudes in HBChPT at ${\cal O}(p^3)$, 
one extracts from Eq.~(\ref{gp}) the following expressions for the GPs
at LO (see Ref.~\cite{HHKD})~:
\begin{small}
\begin{eqnarray}
\left[P^{(M1,L2)1}(\bar{q})\right]^{LO} &=& 
\left[P^{(M1,L0)1}(\bar{q})\right]^{LO} \,=\, \;\;
\left[\hat{P}^{(M1,2)1}(\bar{q})\right]^{LO} \,=\,0, 
\nonumber \\
\left[P^{(M1,M1)1}(\bar{q})\right]^{LO} &=&
\left[P^{(L1,L1)1}(\bar{q})\right]^{LO} \,=\, 0,
\nonumber \\
\left[P^{(L1,M2)1}(\bar{q})\right]^{LO} &=&
\frac{-g_{A}^2}{24\sqrt{2}\pi^{2}F_{\pi}^{2}\bar{q}^{2}}
\left\{ 1-\frac{4}{w\sqrt{w^2+4}}\sinh^{-1}\left(\frac{w}{2}\right) \right\}, 
\nonumber \\
\left[\hat{P}^{(L1,1)1}(\bar{q})\right]^{LO} &=&
\frac{g_{A}^2}{24\sqrt{6}\pi^{2}F_{\pi}^{2}\bar{q}^{2}}
\left\{ 3-\frac{4w^2+12}{w\sqrt{w^2+4}}\sinh^{-1}\left(\frac{w}{2}\right)
\right\},  
\label{result1}
\end{eqnarray}
\end{small}
where $w \equiv \bar{q} / m_{\pi}$, with $m_\pi$ the pion mass.
Furthermore, throughout this paper we use the values~:
$g_A = 1.267$, $F_\pi = 0.0924$~GeV, and $m_\pi = 0.14$~GeV.
\newline
\indent
The VCS amplitudes in HBChPT at ${\cal O}(p^4)$, 
yield the GPs at NLO, from Eq.~(\ref{gp}) (see Ref.~\cite{KV02})~:
\begin{small}
\begin{eqnarray}
\left[P^{(M1,L2)1}(\bar{q})\right]^{NLO} &=&
\frac{-g_{A}^{2}}{12\sqrt{6}\pi^2 F_{\pi}^{2}\bar{q}^{2}}
\left\{ 1-\frac{4}{w\sqrt{w^2+4}}\sinh^{-1}\left(\frac{w}{2}\right)\right\}, 
\nonumber \\
\left[P^{(M1,L0)1}(\bar{q})\right]^{NLO} &=&
\frac{g_{A}^{2}}{12\sqrt{3}\pi^2 F_{\pi}^{2}} \left\{ 2-
\frac{3w^2+8}{w\sqrt{w^2+4}}\sinh^{-1}\left(\frac{w}{2}\right) \right\}, 
\nonumber \\
\left[P^{(M1,M1)1}(\bar{q})\right]^{NLO} &=&
\frac{g_{A}^{2}}{24\pi^2 F_{\pi}^{2}M_{N}}
\left\{ 1-\frac{w^2+4}{w\sqrt{w^2+4}}\sinh^{-1}\left(\frac{w}{2}\right) 
\right\},
\nonumber \\ 
\left[\hat{P}^{(M1,2)1}(\bar{q})\right]^{NLO} &=&
\frac{-g_{A}^{2}}{24\sqrt{10}\pi^2 F_{\pi}^{2}M_{N}\bar{q}^{2}}
\left\{ 3-\frac{2w^2+12}{w\sqrt{w^2+4}}\sinh^{-1}\left(\frac{w}{2}\right)
\right\},
\nonumber \\
\left[P^{(L1,L1)1}(\bar{q})\right]^{NLO} &=& 0, \nonumber \\
\left[P^{(L1,M2)1}(\bar{q})\right]^{NLO} &=&
\frac{g_{A}^{2}}{96\sqrt{2}\pi F_{\pi}^{2} \bar{q}^2} 
\cdot {{\bar q} \over {M_N}} \nonumber \\
&\times & \left\{ \frac{1}{2w}+\frac{2w^2+4}{w(w^2+4)}
+(\frac{5}{4}-\frac{3}{w^2})\tan^{-1}\left(\frac{w}{2}\right) +\tau_{3}
\left[ \frac{1}{2w}+(\frac{1}{4}-\frac{1}{w^2})
\tan^{-1}\left( \frac{w}{2}\right) \right] \right\}, 
\nonumber \\
\left[\hat{P}^{(L1,1)1}(\bar{q})\right]^{NLO} &=&
\frac{-g_{A}^{2}}{96\sqrt{6}\pi F_{\pi}^{2} \bar{q}^2} 
\cdot {{\bar q} \over {M_N}} \nonumber \\
&\times & \left\{ \frac{11}{2w}-\frac{2w^2+4}{w(w^2+4)}
-(\frac{25}{4}+\frac{9}{w^2})\tan^{-1}\left(\frac{w}{2}\right)
+\tau_{3}\left[ \frac{3}{2w}-(\frac{5}{4}+\frac{3}{w^2})
\tan^{-1}\left(\frac{w}{2}\right) \right]\right\} .
\label{eq:nlo}
\end{eqnarray}
\end{small}
\newline
\indent
One can get more predictions by use of the crossing relations~(\ref{con2}), 
which hold in general in a relativistic quantum
field theory, and which were calculated 
in HBChPT to ${\cal O}(p^4)$ in Ref.~\cite{KV02}. 
By plugging in the fourth order amplitudes in (\ref{con2}), we can
extract fifth order predictions from 
${\cal S}_{3}^{(4)}= -2M_{N} / {\bar{q}} \, {\cal S}_{7}^{(5)},$ 
and ${\cal S}_{10}^{(4)}-{\cal S}_{12}^{(4)}=$
$-2M_{N} / {\bar{q}} \, {\cal S}_{11}^{(5)}$. 
In this way, these relations allow us to extract two spin GPs at
next-to-next-to-leading order (NNLO)~: 
\begin{small}
\begin{eqnarray}
\left[P^{(M1,M1)1}(\bar{q})\right]^{NNLO}&=&\frac{-g_{A}^{2}\bar{q}}
{192\pi F_{\pi}^{2}M_{N}^{2}}
\left\{ \frac{3}{w}-(\frac{5}{2}+\frac{6}{w^2})\tan^{-1}\left(\frac{w}{2} \right)
+ \tau_{3}\left[\frac{1}{w}-(\frac{1}{2}+\frac{2}{w^2})\tan^{-1}
\left(\frac{w}{2}\right) \right] \right\}, 
\nonumber \\
\left[P^{(L1,L1)1}(\bar{q})\right]^{NNLO}&=&\frac{g_{A}^{2}}{48\pi^2
F_{\pi}^{2}M_{N}}\left\{ -1+\frac{2w^2+4}{w\sqrt{w^2+4}}
\sinh^{-1}\left(\frac{w}{2}\right) \right\}. 
\label{eq:nnlo}
\end{eqnarray}
\end{small}
\newline
\indent
It is surprising that the third order 
calculation~\cite{HHKD} was able to obtain some of these 
NLO and one NNLO results!
Actually, the NLO results for $P^{(M1,L2)1}$ and $P^{(M1,L0)1}$ 
were obtained in Refs.~\cite{HHKS,HHKD}, by performing the
LEX of the amplitudes $\bar A_i$ to second order in $\omega'$, and, by
isolating two terms at order $\omega'^2$ which depend on
those spin GPs. Furthermore, they used the crossing relations~(\ref{con2}) 
to obtain the NLO results for 
$P^{(M1,M1)1}$ and $\hat P^{(M1,2)1}$, 
as well as the NNLO result for $P^{(L1,L1)1}$.
\newline
\indent
In the present work, we generalize this method and obtain the NNLO result for
$P^{(M1,L2)1}$, $P^{(M1,L0)1}$, and $\hat P^{(M1,2)1}$, 
and even the $N^{3}LO$ result for $P^{(L1,L1)1}$. 
The first step is to define the following quantities:
\begin{equation}
\left[\frac{\partial^{2}\bar{A}_{i}(\omega',\bar{q},\theta)}{\partial
\omega'^{2}}\right]_{\omega'=0}
=2\alpha_{i}(\bar{q})+2\beta_{i}(\bar{q})\cos\theta
+2\gamma_{i}(\bar{q})\cos^{2}\theta.
\label{eq:2ndder}
\end{equation}
Our goal is to express $\alpha_i$, $\beta_i$ and $\gamma_i$
in terms of GPs.
To achieve this goal, one needs to start from the covariant virtual 
Compton scattering tensor.
The VCS amplitudes are obtained as the contraction of the VCS tensor
$M_{\mu\nu}$ with the polarization vectors of the photons, evaluated between
the nucleon spinors in the initial and the final states,
\begin{equation}
{\cal M}_{VCS} = -ie^2\bar{u}(p_{1})\sum_{i=1}^{12}\varepsilon_{1\mu}
\varepsilon_{2\nu}^{*}\rho^{\mu\nu}_{i}
f_{i}(q^2,q\cdot q',q'\cdot P)u(p_{2}),
\label{eq:vcsampl2}
\end{equation}
where $P=p_{1}+p_{2},$ 
$f_{i}$ are the VCS amplitudes, 
and $\rho_{i}^{\mu\nu}$ are the gauge-invariant 
independent tensor structures for VCS, 
as given by \cite{Mainz1}. 
The key point of our method is to connect the
covariant amplitudes $f_{i}$
and the {\it c.m.} amplitudes $A_{i}$ defined in Eq.~(\ref{eq:vcsampl}).
Subsequently, we will perform a low energy expansion of 
the amplitudes $A_{i}$ and sort out the relations 
between $f_i$ and the $\alpha_{i},\beta_{i}$ and $\gamma_{i}$ 
defined through Eq.~(\ref{eq:2ndder}). 
In addition, the 
GPs can also be expressed in terms of the covariant amplitudes $f_{i}$. 
Comparing both results will enable us to express 
the higher order coefficients of the LEX in terms of GPs. 
The advantage of working with the covariant Compton tensors 
is that symmetry properties due to photon crossing 
as well as nucleon crossing combined with charge conjugation, 
as derived in Ref.~\cite{Mainz1} and detailed in the Appendix, 
are immediately manifest.

\section{Calculation of Generalized Polarizabilities 
at NNLO and N$^3$LO in HBChPT }
\label{sec3}

To extract the GPs at NNLO, we start by connecting 
both sets of VCS amplitudes $A_{i}$ and $f_{i}$ 
of Eqs.~(\ref{eq:vcsampl}) and (\ref{eq:vcsampl2}).   
To this goal, one has to expand 
$\varepsilon_{1\mu}\varepsilon_{2\nu}^{*}\rho_{i}^{\mu\nu}$ 
in Eq.~(\ref{eq:vcsampl2}) in $\omega'$
in the {\it c.m.} frame and also expand the non-Born parts of the 
amplitudes $f_{i}$ in $\omega'$: 
\begin{equation}
f_{i}^{non-Born}=\mathring{f}_{i}(\bar{q})+\omega'\cdot
\left[ g_{i}(\bar{q})+h_{i}(\bar{q})\cos\theta \right] +{\cal O}(\omega'^{2}).
\label{eq:lexfi}
\end{equation}
For the new predictions of three NNLO GPs and one N$^3$LO GP 
which we will derive, we only 
need the amplitudes $A_{10}, A_{11}$ and $A_{12}$. 
The expressions for these amplitudes including terms up to order 
$\omega'^2$ can be found, after some algebra, as
\footnote{Note that we omit the global normalization factor 
$\sqrt{\frac{E_N+M_N}{2M_N}}$ on the {\it rhs} of 
Eqs.~(\ref{eq:a10} - \ref{eq:a12}).}~:
\begin{small}
\begin{eqnarray}
\bar A_{10}&=&
\omega'\left[2\omega_{0}\bar{q}\mathring{f}_{4}
-\frac{\omega_{0}^{3}}{2\bar{q}}\mathring{f}_{5}
-\frac{\omega_{0}\bar{q}}{2}\mathring{f}_{7}
-\frac{2\omega_{0}^{2}}{\bar{q}}\mathring{f}_{10}
-2\omega_{0}\bar{q}\mathring{f}_{11}
-\frac{M_N \omega_{0}^{3}}{\bar{q}}\mathring{f}_{12}\right]
\nonumber\\
&+&\omega'^{2}
\left[2\omega_{0}\bar{q}g_{4}-\frac{\omega_{0}^{3}}{2\bar{q}}g_{5}
-\frac{\omega_{0}\bar{q}}{2}g_{7}-\frac{2\omega_{0}^{2}}{\bar{q}}g_{10}
-2\omega_{0}\bar{q}g_{11}-\frac{M_{N}\omega_{0}^{3}}{\bar{q}}g_{12}\right]
\nonumber\\
&+&\omega'^{2}\cos\theta
\left[2\omega_{0}\bar{q}h_{4}-\frac{\omega_{0}^{3}}{2\bar{q}}h_{5}
-\frac{\omega_{0}\bar{q}}{2}h_{7}-\frac{2\omega_{0}^{2}}{\bar{q}}h_{10}
-2\omega_{0}\bar{q}h_{11}-\frac{M_{N}\omega_{0}^{3}}{\bar{q}}h_{12}\right]
\nonumber\\
&+&\omega'^{2}[-\frac{\omega_{0}^{2}}{2M_{N}\bar{q}}\mathring{f}_{1}
+\omega_{0}\bar{q}\mathring{f}_{2}
+\left(2\bar{q}+\frac{\omega_{0}\bar{q}}{M_{N}}\right)\mathring{f}_{4}
+\left(-\frac{\omega_{0}^{3}}{4M_{N}\bar{q}}-\frac{\omega_{0}^{2}}{\bar{q}}
\right)
\mathring{f}_{5}
+\left(-\frac{4M_{N}\omega_{0}^{2}}{\bar{q}}+2\omega_{0}^{2}\cos\theta\right)
\mathring{f}_{6}
+\left(-\frac{\omega_{0}\bar{q}}{4M_{N}}-\frac{\bar{q}}{2}\right)\mathring{f}_
{7}\nonumber\\
&-&\frac{\omega_{0}^2 M_{N}}{\bar{q}}\mathring{f}_{8}
+\omega_{0}\bar{q}\mathring{f}_{9}+\frac{\bar{q}}{M_{N}}\mathring{f}_{10}
+\left(-2\bar{q}-\frac{\omega_{0}\bar{q}}{M_{N}}
-\frac{4\omega_{0}^{2}}{\bar{q}}\right)
\mathring{f}_{11}
+\left(-
\frac{3\omega_{0}^{3}}{2\bar{q}}+\frac{\omega_{0}\bar{q}}{2}\right)
\mathring{f}_{12}]+{\cal O}(\omega'^3),
\label{eq:a10}\\
\bar A_{11}&=&\omega'[(2\bar{q}^{2}-2\omega_{0}\bar{q}\cos\theta)
\mathring{f}_{4}+
\left(-\frac{\omega_{0}^{2}}{2}+\frac{\omega_{0}^{3}}{2\bar{q}}\cos\theta
\right)
\mathring{f}_{5}
+\left(-\frac{\omega_{0}^{2}}{2}+\frac{\omega_{0}\bar{q}}{2}\cos\theta\right)
\mathring{f}_{7}
+\left(-2\omega_{0}+\frac{2\omega_{0}^{2}}{\bar{q}}\cos\theta\right)\mathring{f}_{10}
\nonumber \\
&+&\left(-2\omega_{0}^{2}+2\omega_{0}\bar{q}\cos\theta\right)\mathring{f}_{11}
+\left(-2M_N \omega_{0}^{2}+\frac{M_{N}\omega_{0}^{3}}{\bar{q}}
\cos\theta\right)\mathring{f}_{12}]
\nonumber\\
&+&\omega'^{2}[\left(2\bar{q}^{2}-2\omega_{0}\bar{q}\cos\theta\right)g_{4}
+\left(-\frac{\omega_{0}^{2}}{2}+\frac{\omega_{0}^{3}}{2\bar{q}}\cos\theta
\right)g_{5}
+\left(-\frac{\omega_{0}^{2}}{2}+\frac{\omega_{0}\bar{q}}{2}\cos\theta
\right)g_{7}
+\left(-2\omega_{0}+\frac{2\omega_{0}^{2}}{\bar{q}}\cos\theta\right)
g_{10}\nonumber\\
&+&\left(-2\omega_{0}^{2}+2\omega_{0}\bar{q}\cos\theta\right)g_{11}
+\left(-2M_N\omega_{0}^{2}+\frac{M_{N}\omega_{0}^{3}}{\bar{q}}\cos\theta
\right)g_{12}]
\nonumber\\
&+&\omega'^{2}\cos\theta[\left(2\bar{q}^{2}-2\omega_{0}\bar{q}\cos\theta
\right)h_{4}
+\left(-\frac{\omega_{0}^{2}}{2}+\frac{\omega_{0}^{3}}{2\bar{q}}
\cos\theta\right)h_{5}
+\left(-\frac{\omega_{0}^{2}}{2}+\frac{\omega_{0}\bar{q}}{2}\cos\theta
\right)h_{7}
+\left(-2\omega_{0}+\frac{2\omega_{0}^{2}}{\bar{q}}\cos\theta
\right)h_{10}\nonumber\\
&+&(-2\omega_{0}^{2}+2\omega_{0}\bar{q}\cos\theta)h_{11}
+\left(-2M_{N}\omega_{0}^{2}+\frac{M_{N}\omega_{0}^{3}}{\bar{q}}\cos\theta
\right)h_{12}]
\nonumber\\
&+&\omega'^{2}[-(4M_{N}+2\bar{q}\cos\theta)\mathring{f}_{4}
-\left(\omega_{0}-\frac{\omega_{0}^{2}}{\bar{q}}\cos\theta\right)
\mathring{f}_{5}
+\left(-2\omega_{0}^{2}(1+\cos^2\theta)+\frac{2\omega_{0}}{\bar{q}}\cos\theta(\omega_{0}^2+\bar{q}^2)\right)\mathring{f}_{6}\nonumber \\
&+&\left(-\frac{\omega_{0}}{2}-\frac{M_{N}\omega_{0}}{\bar{q}}\cos\theta\right)
\mathring{f}_{7}
+\left(\omega_{0}^{2}-(\frac{M_{N}\omega_{0}^{2}}{\bar{q}}+\omega_{0}
\bar{q})\cos\theta\right)\mathring{f}_{8}
+\left(\omega_{0}^{2}-\omega_{0}\bar{q}\cos\theta\right)\mathring{f}_{9}
+\left(-2+\frac{2\omega_{0}}{\bar{q}}\cos\theta\right)\mathring{f}_{10}\nonumber \\
&+&\left(-6\omega_{0}+(\frac{2\omega_{0}^{2}}{\bar{q}}+2\bar{q})\cos\theta
\right)
\mathring{f}_{11} 
+\left(-\frac{3\omega_{0}^{2}}{2}-2M_{N}\omega_{0}+\frac{M_{N}
\omega_{0}^{2}}{\bar{q}}\cos\theta\right)\mathring{f}_{12}]
+{\cal O}(\omega'^3),
\label{eq:a11} \\
\bar A_{12}&=&\omega'\left[
-\frac{M_{N}\omega_{0}^{2}}{\bar{q}}
\mathring{f}_{5}+
\left(-\frac{2M_N^{2}\omega_{0}^{2}}{\bar{q}}+M_{N}\omega_{0}\bar{q}\right)
\mathring{f}_{12}\right]\nonumber\\
&+&\omega'^{2}\left[-\frac{M_{N}\omega_{0}^{2}}{\bar{q}}g_{5}+
\left(-\frac{2M_N^{2} \omega_{0}^{2}}{\bar{q}}+M_{N}\omega_{0}\bar{q}\right)
g_{12}\right]
+\omega'^{2}\cos\theta\left[-\frac{M_{N}\omega_{0}^{2}}{\bar{q}}h_{5}+
\left(-\frac{2M_N^{2}\omega_{0}^{2}}{\bar{q}}+M_{N}\omega_{0}\bar{q}\right)
h_{12}\right]\nonumber\\
&+&\omega'^{2}[\left(\frac{\omega_{0}\bar{q}}{M_{N}}
-\frac{\omega_{0}^{2}}{M_{N}}\cos\theta\right)\mathring{f}_{4}
+\left(\frac{\bar{q}}{2}-\frac{3\omega_{0}^{2}}{2\bar{q}}
-\frac{\omega_{0}\bar{q}}{4M_{N}}+\frac{\omega_{0}^{2}}{4M_{N}}
\cos\theta\right)\mathring{f}_{5}
+\left(-\frac{4 M_N \omega_{0}^{2}}{\bar{q}}
+4M_{N}\omega_{0}\cos\theta\right)\mathring{f}_{6}\nonumber \\
&+&\left(-\frac{\omega_{0}\bar{q}}{4M_{N}}+\frac{\bar{q}^{2}}{4M_{N}}
\cos\theta\right)
\mathring{f}_{7}
-\left(\frac{\omega_{0}^{3}}{2\bar{q}}+\frac{\omega_{0}\bar{q}}{2}-\omega_{0}^{2}\cos\theta\right)\mathring{f}_{8}
+\bar{q}(-\omega_{0}+\bar{q}\cos\theta)\mathring{f}_{9}
+\left(\frac{\omega_{0}^{2}}{M_{N}\bar{q}}-\frac{\omega_{0}}{M_{N}}\cos\theta\right)
\mathring{f}_{10}\nonumber \\
&+&\left(-\frac{2\omega_{0}^{2}}{\bar{q}}(1+\frac{\omega_{0}}{2M_{N}})
+(2\omega_{0}+\frac{\omega_{0}^{2}}{M_{N}})\cos\theta\right)\mathring{f}_{11}
+
\left(2M_{N}\bar{q}-\frac{3M_{N}\omega_{0}^{2}}{\bar{q}}+\omega_{0}\bar{q}
+\frac{\omega_{0}^{2}}{2}\cos\theta\right)\mathring{f}_{12}]
\nonumber \\
&+&{\cal O}(\omega'^3). 
\label{eq:a12} 
\end{eqnarray}
\end{small}
We can simplify these expressions by using the crossing symmetry 
relations for the amplitudes $f_i$, as given in the Appendix, 
which lead to  
$\mathring{f}_{4} = \mathring{f}_{8} = \mathring{f}_{10} = 0$.
It is now straightforward to read off from 
Eqs.~(\ref{eq:a10} - \ref{eq:a12}) the coefficients ${\cal S}_{i}$ 
of Eq.~(\ref{eq:linterm}), which 
appear in the LEX of the {\it c.m.} amplitudes $\bar A_i$,  
and express them as combinations of the $\mathring{f}_{i}$~:
\begin{eqnarray}
{\cal S}_{10}&=&
\left[-\frac{\omega_{0}^{3}}{2\bar{q}}\mathring{f}_{5}
-\frac{\omega_{0}\bar{q}}{2}\mathring{f}_{7}
-2\omega_{0}\bar{q}\mathring{f}_{11}
-\frac{M_{N}\omega_{0}^{3}}{\bar{q}}\mathring{f}_{12}\right],
\nonumber\\
{\cal S}_{11}&=&\left[
-\frac{\omega_{0}^{2}}{2} \mathring{f}_{5}
-\frac{\omega_{0}^{2}}{2} \mathring{f}_{7}
-2\omega_{0}^{2} \mathring{f}_{11}
-2M_{N}\omega_{0}^{2} \mathring{f}_{12}\right],
\nonumber\\
{\cal S}_{12}&=&\left[-\frac{M_{N}\omega_{0}^{2}}{\bar{q}}
\mathring{f}_{5}+
\left(-\frac{2M^{2}_{N}\omega_{0}^{2}}{\bar{q}}+M_{N}\omega_{0}\bar{q}\right)
\mathring{f}_{12}\right].
\label{Si}
\end{eqnarray}
By applying the defining relations of Eq.~(\ref{gp}) for the GPs, 
one then obtains~: 
\begin{eqnarray}
P^{(M1,L2)1}(\bar q) &=&\frac{2\sqrt{2}}{3\sqrt{3}}\sqrt{\frac{E_N+
M_N}{2E_N}}
\left[
\frac{\omega_0^2}{2\bar{q}^2}\mathring{f}_{5}+
\frac{1}{2}\mathring{f}_{7}
+2\mathring{f}_{11}+
\frac{M_{N}\omega_0^2}{\bar{q}^2}\mathring{f}_{12}
\right],\nonumber \\
P^{(M1,L0)1}(\bar q) &=&
\frac{2}{\sqrt{3}}\sqrt{\frac{E_N+M_N}{2E_N}}
\left[
\left(
\frac{1}{3}\omega_0^2-M_{N}\omega_0\right)\mathring{f}_{5}
+\frac{1}{3}\bar{q}^{2}\mathring{f}_{7}
+\frac{4}{3}\bar{q}^{2}\mathring{f}_{11}
+\left(M_{N}\bar{q}^{2}-2M^2_{N}\omega_0+\frac{2}{3}M_{N}\omega_0^2\right)
\mathring{f}_{12}
\right],\nonumber \\
P^{(L1,L1)1}(\bar q) &=&
-\frac{2}{3}\sqrt{\frac{E_N+M_N}{2E_N}}
\left[-\frac{\omega_{0}}{2} \mathring{f}_{5}
-\frac{\omega_{0}}{2} \mathring{f}_{7}
-2\omega_{0} \mathring{f}_{11}
-2M_{N}\omega_{0} \mathring{f}_{12}\right].
\label{eq:3gp}
\end{eqnarray}
\newline
\indent
In general, the coefficients of order $\omega'^{2}$ are combinations of
$\mathring{f},\,\,g_{i}$ and $h_{i}$. On the other hand, 
the dipole GPs defined through Eq.~(\ref{gp}) 
are all combinations of $\mathring{f}$. 
Actually $g_{i}$ and $h_{i}$ are related to higher order polarizabilities
$(L'>1)$, and they can be defined in a similar way as in Ref.~\cite{GLT}. 
We leave the study of such higher order GPs to a future work and 
concentrate in this work on new predictions for dipole $(L'=1)$ GPs.
\newline
\indent
To compare with the amplitudes obtained from HBChPT, it is necessary to
make the heavy baryon expansions on Eqs.~(\ref{eq:a10} - \ref{eq:a12}). 
Note that not only $\omega_{0}$ has to be expanded in $\bar q / M_N$ 
as in Eq.~(\ref{eq:omegao}), 
but also $\mathring{f}_{i} ,\,g_{i}$ and $h_{i}$.
Denoting any quantity proportional to $(1/M_N)^n$ by a superscript (n), 
we can express $\mathring{f}_{i}$ as~: 
\begin{equation}
\mathring{f}_{i}=\mathring{f}_{i}^{(0)}+\mathring{f}_{i}^{(1)}
+{\cal O}(1/M^{2}_{N}),\,\,i=1,5,6,7,9,11.
\label{eq:hbchpt1}
\end{equation}
The amplitude $f_{12}$ 
as well as the combination $2 f_6 + f_9$ start with a contribution at $n=1$   
otherwise the amplitudes $\bar{A}_{8}$ and $\bar{A}_{12}$ 
would start at $n=-1$, which is obviously impossible.
Also $f_{2}$ starts at $n=1$ (otherwise $\bar{A}_9$ 
would also start at $n=-1$), 
so that we can write~:
\begin{equation}
\mathring{f}_{i}=\mathring{f}_{i}^{(1)}+{\cal O}(1/M^{2}_{N}),\,\,i=2,12.
\label{eq:hbchpt2}
\end{equation}
The coefficients in the heavy baryon expansion are furthermore 
constrained by the crossing symmetry relations 
for the amplitudes $f_i$ (for details, see the Appendix). 
We can then separate the leading and subleading terms 
in the heavy baryon expansion for the amplitudes 
$\bar A_{10}, \bar A_{11}$, and $\bar A_{12}$ as~:
\begin{small}
\begin{eqnarray}
\bar A_{10}^{LO}&=&
\omega'^{2}\left[-\frac{\bar{q}}{2}\mathring{f}_{7}^{(0)}-2\bar{q}
\mathring{f}_{11}^{(0)}\right],\nonumber \\
\bar A_{11}^{LO}&=&\omega'^{2}[2\bar{q}^{2}g_{4}^{(0)}]
+\omega'^{2}\cos\theta\left[\frac{\bar{q}}{2}\mathring{f}_{7}^{(0)}
+2\bar{q}\mathring{f}_{11}^{(0)}
\right],\nonumber \\
\bar A_{12}^{LO}&=&\omega'^{2}\left[\frac{\bar{q}}{2}\mathring{f}_{5}^{(0)}
+2M_N \bar{q}\mathring{f}_{12}^{(1)}\right]
+\omega'^{2}\cos\theta[-2\bar{q}^{2}\mathring{f}_{6}^{(0)}+
\bar{q}^{2}\mathring{f}_{9}^{(0)}],\nonumber \\
\bar A_{10}^{NLO}&=&\omega'^{2}\left[-\frac{\bar{q}}{2}\mathring{f}_{7}^{(1)}
-2\bar{q}\mathring{f}_{11}^{(1)}\right]
+\frac{1}{M_{N}}\omega'\left[\frac{\bar{q}^{3}}{4}\mathring{f}_{7}^{(0)}
+\bar{q}^{3}\mathring{f}_{11}^{(0)}\right]
+\frac{1}{M_{N}}\omega'^{2}[
-\bar{q}^{3}g_{4}^{(0)}]
+\frac{1}{M_{N}}\omega'^{2}\cos\theta\left[
\frac{\bar{q}^{3}}{4}h_{7}^{(0)}+\bar{q}^{3}h_{11}^{(0)}\right],\nonumber \\
\bar A_{11}^{NLO}&=&\omega'^{2}[
\bar{q}^{2}\mathring{f}_{12}^{(1)} 
+2\bar{q}^{2}g_{4}^{(1)}]
+\omega'^{2}\cos\theta
\left[\frac{\bar{q}}{2}\mathring{f}_{7}^{(1)}+
2\bar{q}\mathring{f}_{11}^{(1)} 
\right]\nonumber \\
&+&\frac{1}{M_{N}}\omega'\cos\theta\left[-
\frac{\bar{q}^{3}}{4}\mathring{f}_{7}^{(0)}
-\bar{q}^{3}\mathring{f}_{11}^{(0)}\right]
+\frac{1}{M_{N}}\omega'^{2}\left[\frac{\bar{q}^{2}}{2}\mathring{f}_{5}^{(0)}
+\frac{\bar{q}^{2}}{4}\mathring{f}_{7}^{(0)}
+3\bar{q}^{2}\mathring{f}_{11}^{(0)}
+\bar{q}^{2}g_{10}^{(0)}\right]\nonumber \\
&+&\frac{1}{M_{N}}\omega'^{2}\cos\theta
\left[-\bar{q}^{3}\mathring{f}_{6}^{(0)}+
\frac{\bar{q}^{3}}{2}\mathring{f}_{9}^{(0)}
+2\bar{q}^{3}g_{4}^{(0)}\right]
+\frac{1}{M_{N}}\omega'^{2}\cos^{2}\theta\left[
\frac{-\bar{q}^{3}}{4}h_{7}^{(0)}
-\bar{q}^{3}h_{11}^{(0)}\right],\nonumber \\
\bar A_{12}^{NLO}&=&\omega'^{2}\left[\frac{\bar{q}}{2}\mathring{f}_{5}^{(1)}
+2M_{N}\bar{q}\mathring{f}_{12}^{(2)}\right]
+\frac{1}{M_{N}}\omega'\left[-\frac{\bar{q}^{3}}{4}\mathring{f}_{5}^{(0)}
-M_{N}\bar{q}^{3}\mathring{f}_{12}^{(1)}\right]
+\frac{1}{M_{N}}\omega'^{2}\left[-\bar{q}^{3}\mathring{f}_{6}^{(0)}+
\frac{\bar{q}^{3}}{2}\mathring{f}_{9}^{(0)} \right]
\nonumber \\
&+&\frac{1}{M_{N}}\omega'^{2}\cos\theta\left[-2M_{N}\bar{q}^{2}
\mathring{f}_{6}^{(1)}
+M_{N}\bar{q}^{2}\mathring{f}_{9}^{(1)}
+\frac{\bar{q}^{2}}{4}\mathring{f}_{7}^{(0)}
-\bar{q}^{2}\mathring{f}_{11}^{(0)}-\frac{\bar{q}^{3}}{4}h_{5}^{(0)}
-M_{N}\bar{q}^{3}h_{12}^{(1)}\right].
\label{AHB}
\end{eqnarray}
\end{small}
\newline
\indent
From Eq.~(\ref{AHB}) one can now identify the coefficients 
$\alpha_{i}$, $\beta_{i}$ and $\gamma_{i}$ appearing in the LEX of 
Eq.~(\ref{eq:2ndder}) as follows~:
\begin{eqnarray}
\alpha_{10}^{LO}&=&-\frac{\bar{q}}{2}\mathring{f}_{7}^{(0)}-2\bar{q}\mathring{f}_{11}^{(0)},\;\;\;
\alpha_{10}^{NLO}=-\frac{\bar{q}}{2}\mathring{f}_{7}^{(1)}-2\bar{q}\mathring{f}_{11}^{(1)}+\frac{1}{M_{N}}[
-\bar{q}^{3}g_{4}^{(0)}],\nonumber \\
\beta_{10}^{LO}&=&0,\;\;\;
\beta_{10}^{NLO}=\frac{1}{M_{N}}\left[
\frac{\bar{q}^{3}}{4}h_{7}^{(0)}+\bar{q}^{3}h_{11}^{(0)}\right],\;\;\;
\gamma_{10}^{LO}=0,
\;\;\;
\gamma_{10}^{NLO}= 0,\nonumber \\
\alpha_{11}^{LO}&=&2\bar{q}^{2}g_{4}^{(0)},\;\;\;
\alpha_{11}^{NLO}=
\bar{q}^{2}\mathring{f}_{12}^{(1)} 
+2\bar{q}^{2}g_{4}^{(1)}
+\frac{1}{M_{N}}\left[\frac{\bar{q}^{2}}{2}\mathring{f}_{5}^{(0)}
+\frac{\bar{q}^{2}}{4}\mathring{f}_{7}^{(0)}
+3\bar{q}^{2}\mathring{f}_{11}^{(0)}
+\bar{q}^{2}g_{10}^{(0)}\right],\nonumber \\
\beta_{11}^{LO}&=&\frac{\bar{q}}{2}\mathring{f}_{7}^{(0)}
+2\bar{q}\mathring{f}_{11}^{(0)},\;\;\;
\beta_{11}^{NLO}=
\frac{\bar{q}}{2}\mathring{f}_{7}^{(1)}+
2\bar{q}\mathring{f}_{11}^{(1)} 
+\frac{1}{M_{N}}\left[-\bar{q}^{3}\mathring{f}_{6}^{(0)}+
\frac{\bar{q}^{3}}{2}\mathring{f}_{9}^{(0)}+
2\bar{q}^{3}g_{4}^{(0)}\right],\nonumber \\
\gamma_{11}^{LO}&=&0
,\;\;\;
\gamma_{11}^{NLO}=\frac{1}{M_{N}}\left[
\frac{-\bar{q}^{3}}{4}h_{7}^{(0)}
-\bar{q}^{3}h_{11}^{(0)}\right],\nonumber \\
\alpha_{12}^{LO}&=&\frac{\bar{q}}{2}\mathring{f}_{5}^{(0)}
+2M_{N}\bar{q}\mathring{f}_{12}^{(1)},\;\;\;
\alpha_{12}^{NLO}=\frac{\bar{q}}{2}\mathring{f}_{5}^{(1)}
+2M_{N}\bar{q}\mathring{f}_{12}^{(2)}
-\frac{\bar{q}^{3}}{M_N}\mathring{f}_{6}^{(0)}+
\frac{\bar{q}^{3}}{2M_N}\mathring{f}_{9}^{(0)},
\nonumber \\
\beta_{12}^{LO}&=&[-2\bar{q}^{2}\mathring{f}_{6}^{(0)}+
\bar{q}^{2}\mathring{f}_{9}^{(0)}]\;\;\;
\beta_{12}^{NLO}=\frac{1}{M_{N}}\left[-2M_{N}\bar{q}^{2}\mathring{f}_{6}^{(1)}
+M_{N}\bar{q}^{2}\mathring{f}_{9}^{(1)}
+\frac{\bar{q}^{2}}{4}\mathring{f}_{7}^{(0)}
-\bar{q}^{2}\mathring{f}_{11}^{(0)}-\frac{\bar{q}^{3}}{4}h_{5}^{(0)}
-M_{N}\bar{q}^{3}h_{12}^{(1)}\right], \nonumber \\
\gamma_{12}^{LO}&=&0, \;\;\;
\gamma_{12}^{NLO}=0. 
\label{co}
\end{eqnarray}
\newline
\indent
The GPs of Eq.~(\ref{eq:3gp}) can now be expressed in terms of the above 
LEX coefficients. 
The power counting of the GPs needs some word of explanation though. 
From HBChPT, the leading order amplitudes ${\cal S}_{i}^{(3)}\sim {\cal O}(
1)$ and ${\cal S}_{i}^{(4)}\sim {\cal O}(1/M_{N})$. Because $\omega_{0}$
is a quantity of order ${\cal O}(1/M_{N})$, one remarks from Eq.~(\ref{gp}) 
that the three GPs $P^{(M1,L2)1},\,P^{(L1,L1)1}$ and $P^{(M1,L0)1}$ 
should start at ${\cal O}(M_{N})$. 
Actually ${\cal S}_{10}^{(3)}={\cal S}_{11}^{(3)}
={\cal S}_{12}^{(3)}=0$, and consequently these three GPs are non-zero only 
at NLO~: 
\begin{eqnarray}
\left[P^{(M1,L2)1}(\bar q)\right]^{NLO}&=&\frac{2\sqrt{2}}{3\sqrt{3}}
\left[\frac{1}{2}\mathring{f}_{7}^{(0)}
+2\mathring{f}_{11}^{(0)}\right]
=-\frac{2\sqrt{2}}{3\sqrt{3}}\left[\frac{1}{\bar{q}}\alpha_{10}^{LO} \right],
\nonumber \\
\left[P^{(M1,L0)1}(\bar q)\right]^{NLO}&=&
\frac{2}{\sqrt{3}}\left[\frac{\bar{q}^{2}}{2}\mathring{f}_{5}^{(0)}
+\frac{1}{3}\bar{q}^{2}\mathring{f}_{7}^{(0)}+
\frac{4}{3}
\bar{q}^{2}\mathring{f}_{11}^{(0)}+2M_{N}\bar{q}^{2}\mathring{f}_{12}^{(1)}
\right]
=\frac{2}{\sqrt{3}}\left[\bar{q}\alpha_{12}^{LO}
-\frac{2}{3}\bar{q}\alpha_{10}^{LO}\right], \nonumber \\
\left[P^{(M1,L2)1}(\bar q)\right]^{NNLO}&=&\frac{2\sqrt{2}}{3\sqrt{3}}
\left[\frac{1}{2}\mathring{f}_{7}^{(1)}
+2\mathring{f}_{11}^{(1)}\right]
=\frac{2\sqrt{2}}{3\sqrt{3}}\left[-\frac{1}{\bar{q}}\alpha_{10}^{NLO}
-\frac{1}{2M_N}\alpha_{11}^{LO}\right],
\nonumber \\
\left[P^{(M1,L0)1}(\bar q)\right]^{NNLO}&=&
\frac{2}{\sqrt{3}}\left[\frac{\bar{q}^{2}}{2}\mathring{f}_{5}^{(1)}
+\frac{1}{3}\bar{q}^{2}\mathring{f}_{7}^{(1)}+
\frac{4}{3}
\bar{q}^{2}\mathring{f}_{11}^{(1)}+2M_{N}\bar{q}^{2}\mathring{f}_{12}^{(2)}
\right]\nonumber \\
&=&\frac{2}{\sqrt{3}}\left[
\bar{q}\alpha_{12}^{NLO}
-\frac{\bar{q}^2}{2M_N}\beta_{12}^{LO}
-\frac{2}{3}\bar{q}\alpha_{10}^{NLO}
-\frac{1}{3}\frac{\bar{q}^2}{M_{N}}\alpha_{11}^{LO}
\right].
\label{main}
\end{eqnarray}
\newline
\indent
To evaluate the GPs $P^{(M1,L2)1}$ and $P^{(M1,L0)1}$ at NNLO, 
requires the knowledge of the VCS amplitudes 
at ${\cal O}(p^{5})$ in HBChPT through the coefficients 
$\alpha_{10}^{NLO}$ and $\alpha_{12}^{NLO}$. 
In general, the amplitudes at ${\cal O}(p^{5})$ in HBChPT
contain two parts. The first one 
consists of two-loop diagrams and one-loop diagrams with one vertex
from the third-order Lagrangian ${\cal L}^{(3)}_{\pi N}$. 
The second one consists of 
one-loop diagrams with vertices from the heavy baryon expansion.
The main difference between both contributions is that the former 
ones contain no $1 / M_{N}$ factor whereas the latter ones 
contain a $1 / M_{N}^{2}$ pre-factor. 
We will present here the results due to the one-loop diagrams with vertices 
from the heavy baryon expansion. Taking into account these contributions  
will precisely restore the symmetry property due to nucleon crossing 
combined with charge conjugation, which relates the  
VCS amplitudes at different orders in the heavy baryon expansion. 
The new analytic structures which one can expect to arise 
at ${\cal O}(p^{5})$ due to two-loop effects have then 
to satisfy the crossing relations separately. 
\newline
\indent
We find for $\alpha_{10}^{NLO}$ and $\alpha_{12}^{NLO}$ 
by explicit calculation in HBChPT the following results~:
\begin{eqnarray}
\alpha_{11}^{LO}&=&0,\;\;\;\beta_{12}^{LO}=0, 
\nonumber\\
\alpha_{10}^{NLO}&=&\frac{g_{A}^{2}}{64M_{N}\pi F_{\pi}^{2}}\left\{
\left[\frac{-2}{w}+\frac{4-6w^2}{4w+w^3}-\frac{4}{3}\cdot
\frac{6+15w^2+2w^4}{4w+w^3}
+(12+\frac{6}{w^2})\tan^{-1}\left(\frac{w}{2}\right)\right]
\right. \nonumber \\
&+&\left. \frac{1}{2}[(1+\kappa_{v})
-(1+\kappa_{s})\tau_{3}]\left[
-\frac{8}{w}+ \frac{w}{3}+(2+\frac{16}{w^2})\tan^{-1}\left(\frac{w}{2}\right)
\right]\right\}, \nonumber \\
\alpha_{12}^{NLO}&=&\frac{g_{A}^{2}}{64M_{N}\pi F_{\pi}^{2}}\left\{
\left[\frac{-7}{w}+\frac{2w}{3}+(2+\frac{14}{w^2})\tan^{-1}
\left(\frac{w}{2}\right)\right] \right. \nonumber \\
&+& \frac{1}{2} [1+\kappa_{v}] \left[\frac{-3}{w} + \frac{w}{6}
+3 \, (\frac{1}{2}+\frac{2}{w^2})\tan^{-1}\left(\frac{w}{2}\right)\right]
\nonumber \\
&+& \left. \frac{1}{2} [(1+\kappa_{s})\tau_{3}] 
\left[\frac{-1}{w}-\frac{w}{6}
+(\frac{5}{2}+\frac{2}{w^2})\tan^{-1}\left(\frac{w}{2}\right)\right] \right\},
\label{eq:a10a12}
\end{eqnarray}
where $\kappa_v = 3.70$ ( $\kappa_s = -0.12$ ) are the 
nucleon isovector (isoscalar) anomalous magnetic moments respectively.  
Combining Eqs.~(\ref{main}) and (\ref{eq:a10a12}), allows us to 
obtain the NNLO results for two GPs~:
\begin{eqnarray}
\left[P^{(M1,L0)1}(\bar q)\right]^{NNLO}&=& 
\frac{g_{A}^{2}\bar{q}}{96\sqrt{3}M_{N}\pi F_{\pi}^{2}}\left\{
\left[\frac{-17}{w}+2w-
\frac{8-12w^2}{4w+w^3}+\frac{8}{3}\cdot
\frac{6+15w^2+2w^4}{4w+w^3}
+(-18+\frac{30}{w^2})\tan^{-1}\left(\frac{w}{2}\right)\right] 
\right. \nonumber \\
&+&[1+\kappa_{v}]\left[
\frac{7}{2w}-\frac{w}{12}+(\frac{1}{4}-\frac{7}{w^2})
\tan^{-1}\left(\frac{w}{2}\right)\right] \nonumber \\
&+&\left. [(1+\kappa_{s})\tau_{3}]\left[
-\frac{19}{2w}+\frac{w}{12}+(\frac{23}{4}+\frac{19}{w^2})\tan^{-1}
\left(\frac{w}{2}\right)\right]\right\},
\label{eq:nnlogp1} \\
\left[P^{(M1,L2)1}(\bar q)\right]^{NNLO}&=&
\frac{-g_{A}^{2}}{48\sqrt{6}M_{N}\pi F_{\pi}^{2}\bar{q}}
\left\{
\left[\frac{-2}{w}+\frac{4-6w^2}{4w+w^3}-\frac{4}{3}\cdot
\frac{6+15w^2+2w^4}{4w+w^3}
+(12+\frac{6}{w^2})\tan^{-1}\left(\frac{w}{2}\right)\right]
\right. \nonumber \\
&+&\left.[(1+\kappa_{v})
-(1+\kappa_{s})\tau_{3}]\left[
-\frac{4}{w} + \frac{w}{6}
+(1+\frac{8}{w^2})\tan^{-1}\left(\frac{w}{2}\right)\right] \right\}.
\label{eq:nnlogp2}
\end{eqnarray}
A third NNLO prediction follows by applying the crossing symmetry constraint 
of Eq.~(\ref{con2}), i.e. 
${\cal S}_{10}-{\cal S}_{12}=\frac{\bar{q}}{\omega_{0}} {\cal S}_{11}$, 
which yields the relation between 3 spin GPs~:
\begin{equation}
3\frac{\bar{q}^2}{\omega_{0}}P^{(L1,L1)1}=\sqrt{3}P^{(M1,L0)1}+\sqrt{\frac{3}{2}}\bar{q}^2P^{(M1,L2)1}.
\end{equation}
After the heavy baryon expansion, this becomes~:
\begin{equation}
- 6M_{N} \, [P^{(L1,L1)1}]^{N^3 LO}=
\sqrt{3} [P^{(M1,L0)1}]^{NNLO}
+\sqrt{\frac{3}{2}}\bar{q}^2 [P^{(M1,L2)1}]^{NNLO},
\end{equation}
from which we obtain~:
\begin{eqnarray}
\left[P^{(L1,L1)1}(\bar q)\right]^{N^3LO}&=&
\frac{g_{A}^{2}\bar{q}}{576M_{N}^{2}\pi F_{\pi}^{2}}\left\{
\left[\frac{15}{w}-2w+3\cdot
\frac{4-6w^2}{4w+w^3}-4\cdot
\frac{6+15w^2+2w^4}{4w+w^3}
+(30-\frac{24}{w^2})\tan^{-1}\left(\frac{w}{2}\right)\right]
\right. \nonumber \\
&+&[1+\kappa_{v}]\left[
-\frac{15}{2w} + \frac{w}{4} 
+(\frac{3}{4}+\frac{15}{w^2})\tan^{-1}\left(\frac{w}{2}\right)
\right] \nonumber \\
&+&\left. [(1+\kappa_{s})\tau_{3}]\left[
+\frac{27}{2w} - \frac{w}{4} +(-\frac{27}{4}-\frac{27}{w^2})
\tan^{-1}\left(\frac{w}{2}\right)\right]\right\}.
\label{eq:nnnlogp}
\end{eqnarray} 
Similarly,
by applying the crossing symmetry constraint 
of Eq.~(\ref{con2}), i.e. ${\cal S}_{4}=0$, 
which yields the relation between 3 spin GPs~:
\begin{equation}
\sqrt{\frac{3}{2}}\omega_{0}P^{(M1,L2)1}(\bar{q})+P^{(M1,M1)1}(\bar{q})
+\sqrt{\frac{5}{2}}\bar{q}^2\hat{P}^{(M1,2)1}=0,
\end{equation}
one can get the NNLO result of $\hat{P}^{(M1,2)1}$~:
\begin{eqnarray}
&&\left[\hat{P}^{(M1,2)1}(\bar q)\right]^{NNLO} =
\frac{-g_{A}^{2}}{96\sqrt{10}M_{N}^{2}\pi F_{\pi}^{2}\bar{q}}
\left\{
\left[\frac{-5}{w}+\frac{4-6w^2}{4w+w^3}-\frac{4}{3}\cdot
\frac{6+15w^2+2w^4}{4w+w^3}
+(\frac{29}{2}+\frac{12}{w^2})\tan^{-1}\left(\frac{w}{2}\right)\right]
\right. \nonumber \\
&&\hspace{1.25cm} +\left.[(1+\kappa_{v})
-(1+\kappa_{s})\tau_{3}]\left[
-\frac{4}{w} + \frac{w}{6}
+(1+\frac{8}{w^2})\tan^{-1}\left(\frac{w}{2}\right)\right]
+\tau_{3}\left[
-\frac{1}{w}
+(\frac{1}{2}+\frac{2}{w^2})\tan^{-1}\left(\frac{w}{2}\right)\right]
\right\}.
\label{eq:nnlogp3}
\end{eqnarray}
At the real photon point ( $\bar q = 0$), we can express the 
GP $P^{(M1,L2)1}(0)$ in terms of a sum of two spin polarizabilities 
( $\gamma_2 + \gamma_4$ ), introduced by Ragusa \cite{Rag94}, as~:
$$\gamma_{2}+\gamma_{4}=-\alpha_{em}\frac{3\sqrt{3}}{2\sqrt{2}}
P^{(M1,L2)1}(0),$$
where $\alpha_{em} \equiv e^2 / (4 \pi) = 1/137$. 
The result of Eq.~(\ref{eq:nnlogp2}) then allows to extract the result for 
$\gamma_2 + \gamma_4$ at NLO as~: 
\begin{equation}
(\gamma_{2}+\gamma_{4})^{NLO}=\frac{\alpha_{em}g_{A}^{2}}{384\pi 
F_{\pi}^{2}m_{\pi}M_N}\left[-3+2(1+\kappa_{v})
-2(1+\kappa_{s})\tau_{3}\right], 
\end{equation}
which exactly reproduces the NLO calculations for $\gamma_2 + \gamma_4$ 
of Refs.~\cite{Kum00,Gel00}.
\newline
\indent
From our results, it is also worth noting that the 
crossing relations of Eq.~(\ref{con2}) 
allow to classify the GPs into two groups~:
\begin{eqnarray}
{\mathrm{group}} \, 1 
&=&\left\{P^{(M1,L2)1}, P^{(M1,L0)1}, P^{(M1,M1)1}, 
\hat{P}^{(M1,2)1}\right\}, \nonumber\\
{\mathrm{group}} \, 2 
&=&\left\{P^{(L1,L1)1}, {P}^{(L1,M2)1}, \hat{P}^{(L1,1)1}\right\}. 
\nonumber
\end{eqnarray}
It is interesting to observe that their analytical 
forms alternate from one order to the next, as noticed already 
in Ref.~\cite{KV02}. At LO one has~:
\begin{eqnarray}  
{\mathrm{group}} \,1 :
\frac{1}{\pi}\tan^{-1}\left(\frac{\bar{q}}{2m_{\pi}}\right);
\hspace{0.25cm} {\mathrm{group}} \,2 :
\frac{1}{\pi^2}\sinh^{-1}\left(\frac{\bar{q}}{2m_{\pi}}\right).
\nonumber
\end{eqnarray}
At NLO, one has~:
\begin{eqnarray}  
{\mathrm{group}} \,1 :
\frac{1}{\pi^2}\sinh^{-1}\left(\frac{\bar{q}}{2m_{\pi}}\right);
\hspace{0.25cm} {\mathrm{group}} \,2 :
\frac{1}{\pi}\tan^{-1}\left( \frac{\bar{q}}{2m_{\pi}}\right) .
\nonumber
\end{eqnarray}  
At NNLO one again has the same analytical structure as at LO, etc.
One therefore observes that the alternating analytical forms 
teach us how the crossing relations~(\ref{con2}), 
connecting the GPs between two different orders in
HBChPT, work. The same alternating analytical form 
were observed when comparing leading and next-to-leading order 
HBChPT results for the forward spin polarizabilities 
in Ref.~\cite{KSV03}.

\section{Results and discussion}
\label{sec4}

The main result of this work are the new NNLO predictions of 
Eqs.~(\ref{eq:nnlogp1}, \ref{eq:nnlogp2}) and (\ref{eq:nnlogp3}) for the spin 
GPs $P^{(M1,L0)1}$, $P^{(M1,L2)1}$, and $\hat P^{(M1,2)1}$, as well as the 
N$^3$LO prediction of Eq.~(\ref{eq:nnnlogp}) for the spin 
GP $P^{(L1,L1)1}$. 
\begin{figure}[h]
\includegraphics[width=10cm]{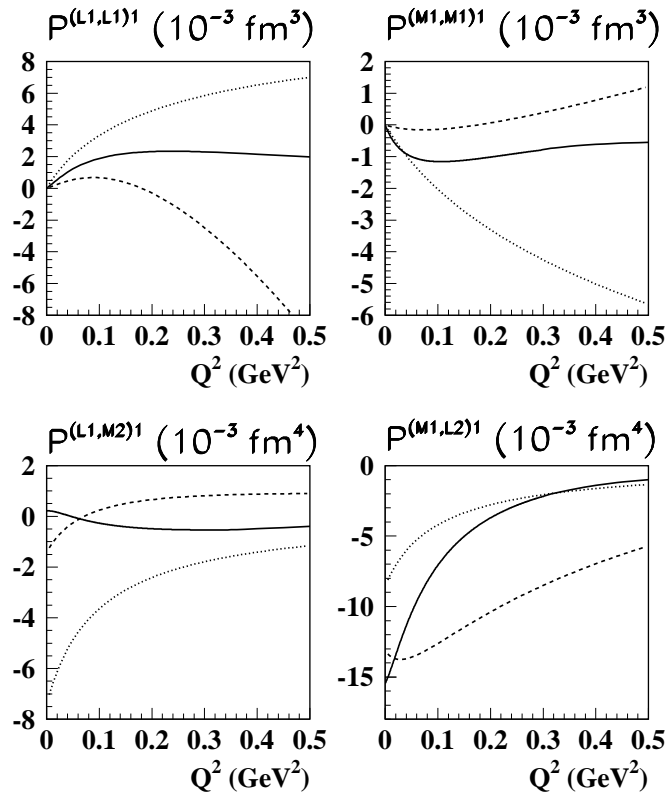}
\caption[]{Results for the spin GPs in HBChPT.
For the GP $P^{(L 1,M 2)1}$, both the LO (dotted curve) and 
NLO (dashed curve) results are shown. 
For the GPs $P^{(M 1,M 1)1}$ and $P^{(M1,L2)1}$, 
both the NLO results (dotted curves) and 
the NNLO results (dashed curves) are shown.
For the GP $P^{(L 1,L 1)1}$, the NNLO (dotted curve) and 
N$^3$LO (dashed curve) results are shown.
For comparison, we also show the 
dispersive evaluation of Refs.~\cite{Pas00,Pas01} (solid curves).
}
\label{fig:polarizab_comp}
\end{figure}
\newline
\indent
In Fig.~\ref{fig:polarizab_comp}, we compare these results 
for the GPs $P^{(M1,L2)1}$ and $P^{(L1,L1)1}$ 
with their lowest order non-zero expressions, 
and display as well the next order results 
for the GPs $P^{(L1,M2)1}$ and $P^{(M1,M1)1}$, 
which were calculated before in Ref.~\cite{KV02}. 
Note that the three remaining spin GPs of Eq.~(\ref{gp}) 
can be expressed in terms of these four independent spin GPs by 
making use of the crossing symmetry relations of Eq.~(\ref{con2}). 
By comparing the lowest order with next order results in HBChPT 
for each of the GPs, one observes large corrections at next order. 
We also compare these results in Fig.~\ref{fig:polarizab_comp} with the 
phenomenological estimate of Refs.~\cite{Pas00,Pas01} using 
dispersion relations. 
One sees that for the GPs $P^{(L1,M2)1}, P^{(M1,M1)1}$ and $P^{(L1,L1)1}$ 
the corrections to the leading order bring the HBChPT results 
towards the phenomenological estimates, though overshoots it for 
$P^{(L1,L1)1}$. For the GP $P^{(M1,L2)1}$, the sizeable large correction 
at the real photon point brings the HBChPT results close to the dispersive 
results but shows a much slower fall-off with $Q^2$ 
than the dispersive estimate.  
The comparison in Fig.~\ref{fig:polarizab_comp} clearly indicates that a 
satisfactory description of spin GPs is still a challenging task. 
To completely disentangle the four spin GPs experimentally, requires four 
independent observables, which we discuss in the following.  
\newline
\indent 
It was shown in Ref.~\cite{GLT} that GPs can be accessed 
experimentally by 
measuring the $e p \to e p \gamma$ reaction and performing a 
low energy expansion in the outgoing photon energy $\omega'$. 
In particular, the VCS unpolarized 
squared amplitude $\calm^{\rm exp}$ takes on the form \cite{GLT}~:
\begin{equation}
\calm^{\rm exp}=\frac{\calm^{\rm exp}_{-2}}{\omega'^2}
+\frac{\calm^{\rm exp}_{-1}}{\omega'}
+\calm^{\rm exp}_0+O(\omega') \, .
\label{eq:unpolsqramp}
\end{equation}
Due to the low energy theorem (LET), the threshold coefficients 
$\calm^{\rm exp}_{-2}$ and $\calm^{\rm exp}_{-1}$ are
known~\cite{GLT}, and are fully determined from the Bethe-Heitler +
Born (BH + B) amplitudes. 
The information on the GPs is contained in \( \calm^{\rm exp}_0\), 
which contains a part originating from the 
BH+B amplitudes and another one which is a linear combination of the GPs, with
coefficients determined by the kinematics. 
The unpolarized observable $\calm^{\rm exp}_0$ 
can be expressed in terms of three structure functions
$P_{LL}(\bar q)$, $P_{TT}(\bar q)$, and $P_{LT}(\bar q)$ as~\cite{GLT}~:
\begin{eqnarray}
\hspace{-0.7cm}
\calm^{\rm exp}_0 - \calm^{\rm BH+B}_0 
= 2 K_2  \Bigg\{ 
v_1 \left[ {\varepsilon  P_{LL}(\bar q) - P_{TT}}(\bar q)\right] 
+ \left(v_2-\frac{\omega_0}{\bar q}v_3\right)\sqrt {2\varepsilon \left( 
{1+\varepsilon }\right)} P_{LT}(\bar q) \Bigg\}, 
\label{eq:vcsunpol}
\end{eqnarray}
where $K_2$ is a kinematical factor, $\varepsilon$ is the virtual
photon polarization (in the standard notation used in electron
scattering), and $v_1, v_2, v_3$ are kinematical
quantities depending on $\varepsilon$ and $\bar q$  
as well as on the {\it c.m.} polar and azimuthal angles
($\theta_{\gamma\gamma}$  and $\phi$, respectively) of  the
produced real photon (for details see Ref.~\cite{GV98}). 
The three unpolarized observables of Eq.~(\ref{eq:vcsunpol}) 
can be expressed in terms of the independent GPs as \cite{GLT,GV98}~:
\begin{eqnarray}
&&P_{LL} \,=\, - 2\sqrt{6} \, M_N \, G_E(Q^2) 
\, P^{\left( {L1,L1} \right)0} \;, 
\label{eq:unpolobsgp1} \\
&&P_{TT} \,=\, - 3 \, G_M(Q^2) \, \frac{\bar q^2}{\omega_0} 
\left( P^{(M1,M1)1}\,-\,\sqrt{2} \, \omega_0 \, P^{(L1,M2)1} \right) , 
\label{eq:unpolobsgp2} \\
&&P_{LT} \,=\, \sqrt{\frac{3}{2}} \, \frac{M_N \, \bar q}{Q} \, G_E(Q^2) \, 
P^{(M1,M1)0} \,+\,\frac{3}{2} \, \frac{Q \, \bar q}{\omega_0}\,G_M(Q^2) \,
P^{(L1,L1)1}, 
\label{eq:unpolobsgp3}
\end{eqnarray}
where $Q^2 \equiv \bar q^2 - \omega_0^2 = - 2 M_N \omega_0$, and 
$G_E(Q^2)$ and $G_M(Q^2)$ stand for the electric and magnetic 
nucleon form factors respectively. 
The spin independent GPs $P^{(L1,L1)0}$ and $P^{(M1,M1)0}$ are directly 
proportional to the electric and magnetic polarizabilities of the 
nucleon respectively. 
A Fourier transform of their $\bar q$-dependence allows us to 
map out the spatial distribution of the electric polarization and 
magnetization of the nucleon, as discussed in Ref.~\cite{spatial}.
\newline
\indent
The first VCS experiment was performed at MAMI \cite{Roc00} and 
the response functions $P_{LT}$ and $P_{LL}$~-~$P_{TT}/\varepsilon$ 
were extracted at $Q^2$ = 0.33 GeV$^2$. 
Going to higher $Q^2$, the VCS process has also been measured at JLab 
and data have been obtained  
at $Q^2$ = 1 GeV$^2$ and $Q^2$ = 1.9 GeV$^2$ \cite{Lav04}. 
To unambiguously extract the spin-independent GPs from experiment, 
requires to separate $P_{LL}$ from $P_{TT}$. 
This can be done by performing two unpolarized VCS experiments 
at a fixed value of $\bar q$ (or equivalently $Q^2$) 
and by varying $\varepsilon$ (or equivalently the beam energy). 
Such experiments are planned at MAMI in the near future, making use 
of the beam energy upgrade. 
Besides extracting the electric GP $P^{(L1,L1)0}$, such an 
experiment will also allow for a measurement of the 
combination of the two spin GPs 
$P^{(M1,M1)1}$ and $P^{(L1,M2)1}$ of Eq.~(\ref{eq:unpolobsgp2}).
\begin{figure}[h]
\includegraphics[width=11cm]{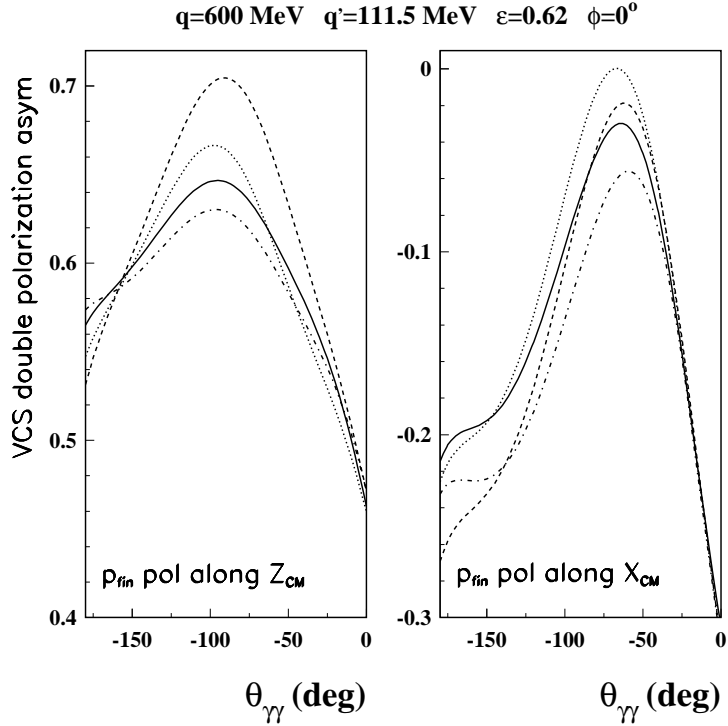}
\caption[]{VCS double-polarization asymmetry (polarized electron,
recoil proton polarization along either the $z$- or $x$- directions 
in the {\it c.m.} frame) 
in MAMI kinematics as function of the photon scattering angle.
The dashed-dotted curves correspond to the BH+B contribution.
The dotted curves are the lowest order HBChPT predictions 
from Refs.~\cite{HHKS,HHKD}.
The dashed curves are obtained by including the next order HBChPT 
correction for all spin-flip GPs, as calculated in this work, 
while using the lowest order predictions for the spin-independent GPs.
For comparison, we also show the results of a phenomenological dispersion 
relation calculation \cite{Pas01,Dre03} (solid curves).}
\label{fig:fig2}
\end{figure}
\newline
\indent
Until now, we discussed only unpolarized VCS observables. 
An unpolarized VCS experiment gives access to only three combinations of
the six GPs, as given by Eqs.~(\ref{eq:unpolobsgp1}) - (\ref{eq:unpolobsgp3}). 
It was shown in Ref.~\cite{Vdh97} that VCS double
polarization observables with polarized lepton beam and polarized target
(or recoil) nucleon,
will allow us to measure three more combinations of 
spin-flip GPs. Therefore a
measurement of unpolarized VCS observables (at different values of 
$\varepsilon$) and of three double-polarization observables 
will enable us to disentangle all six GPs. 
The VCS double polarization observables, which are denoted by 
\(\Delta \calm (h, i) \) for an electron of helicity $h$, are defined as 
the difference of the squared amplitudes for recoil (or target) proton
spin orientation in the direction and opposite to 
the axis $i$ ($i = x, y, z$), where 
the $z$-direction is chosen along the virtual photon momentum 
(see Ref.~\cite{Vdh97} for details). 
In a LEX, this polarized squared amplitude yields~:   
\begin{equation}
\dcalm^{\rm exp}=\frac{\dcalm^{\rm exp}_{-2}}{\omega'^2}
+\frac{\dcalm^{\rm exp}_{-1}}{\omega'}+\dcalm^{\rm exp}_0+O(\omega') \,. \;\;
\label{eq_3_50}
\end{equation}
Analogously to the unpolarized squared amplitude 
of Eq.~(\ref{eq:unpolsqramp}), 
the threshold coefficients $\dcalm^{\rm exp}_{-2}$, 
$\dcalm^{\rm exp}_{-1}$ are known due
to the LET. 
It was found in Ref.~\cite{Vdh97} that 
the polarized squared amplitude $\Delta \calm^{\rm exp}_0$ 
can be expressed in terms of three new structure functions
$P_{LT}^z(\bar q)$, $P_{LT}^{'z}(\bar q)$, and $P_{LT}^{'\perp}(\bar q)$. 
These new structure functions are related to the spin GPs 
as~\cite{Vdh97,GV98}~: 
\begin{eqnarray}
&&P_{LT}^z \;=\; \frac{3 \, Q \,\bar q}{2 \, \omega_0} \, G_M(Q^2) 
\, P^{(L1,L1)1}
\,-\, \frac{3\, M_N \, \bar q}{Q} \, G_E(Q^2) \, P^{(M1,M1)1}, 
\label{eq:polobsgp1} \\
&&P_{LT}^{'z} \;=\; -\frac{3}{2}\, Q \, G_M(Q^2) \, P^{(L1,L1)1}
\,+\, \frac{3 \, M_N \, \bar q^2}{Q \, \omega_0} \, G_E(Q^2) \, P^{(M1,M1)1}, 
\label{eq:polobsgp2} \\
&&P_{LT}^{'\perp} \;=\; \frac{3 \,\bar q \, Q}{2 \, \omega_0} \,G_M(Q^2) \, 
\left(P^{(L1,L1)1} \,-\, \sqrt{\frac{3}{2}} \, \omega_0 \, P^{(M1,L2)1}\right).
\hspace{.4cm} 
\label{eq:polobsgp3}
\end{eqnarray}
While $P_{LT}^z$ and $P_{LT}^{'z}$ can be accessed by in-plane
kinematics ($\phi = 0^{\mathrm{o}}$), the measurement of $P_{LT}^{'\perp}$
requires an out-of-plane experiment. 
\newline
\indent
In Fig.~\ref{fig:fig2}, 
we show the results for the double polarization
observables, with polarized electron and by measuring the recoil
proton polarization either along the virtual photon direction
($z$-direction) or parallel to the reaction plane and perpendicular to
the virtual photon ($x$-direction). 
These asymmetries directly depend on the response functions 
of Eqs.~(\ref{eq:polobsgp1})-(\ref{eq:polobsgp3}).
One sees from Fig.~\ref{fig:fig2} that the double polarization
asymmetries are quite large (due to a non-vanishing
asymmetry for the BH + B mechanism).
We furthermore compare in Fig.~\ref{fig:fig2} the HBChPT predictions 
for the lowest order (non-vanishing) results for the spin GPs 
from Refs.~\cite{HHKS,HHKD}, with the next-order results for the spin GPs, 
as calculated in this work. For the spin independent GPs, which also enter 
in these asymmetries, we use in both cases the LO results, 
awaiting a NLO calculation for $P^{(L1,L1)0}$ and $P^{(M1,M1)0}$. 
One observes from Fig.~\ref{fig:fig2} sizeable differences 
in the asymmetries between the leading order 
and next order HBChPT predictions for 
the spin GPs. For comparison, we also show in Fig.~\ref{fig:fig2} the 
results of the DR calculation of Refs.~\cite{Pas01,Dre03}. 
Due to the smaller spin GPs in the dispersion calculation, the 
relative effect on the asymmetries is smaller in the DR approach. 
\newline
\indent
To quantify the relative effect on the asymmetries due to the spin GPs, we 
display in Fig.~\ref{fig:fig3} the double polarization asymmetries with 
the well known Bethe-Heitler + Born contribution subtracted. 
Another contribution which is well known is the anomaly contribution 
(due to a $t$-channel $\pi^0$ pole) to the spin GPs, which is shown 
in Fig.~\ref{fig:fig3} (lower panels). 
One furthermore sees from Fig.~\ref{fig:fig3} (upper panels) 
that the effect of the spin GPs, excluding the anomaly contribution, 
on the double polarization asymmetries is of the order of 5 \%. 
To distinguish between the different predictions, 
a double polarization VCS experiment is clearly called for. 
Although these double polarization observables are tough to measure, 
a first test experiment is already underway at MAMI \cite{dHo01}.
\begin{figure}[h]
\includegraphics[width=11cm]{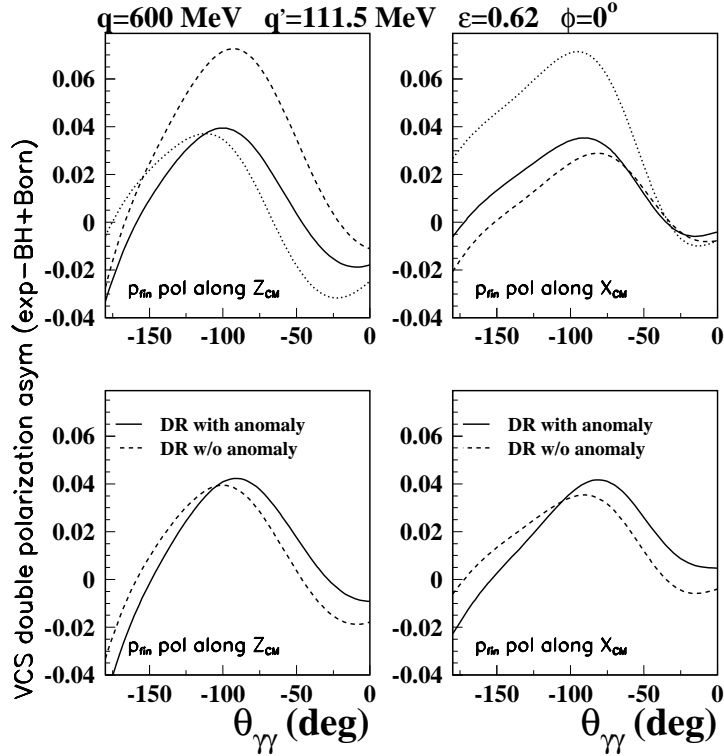}
\caption[]{Deviation of the double-polarization VCS asymmetry 
from the BH+Born result, calculated within the LEX formalism 
in MAMI kinematics as function of the photon scattering angle.
Upper panels: lowest order HBChPT predictions from Refs.~\cite{HHKS,HHKD} 
(dotted curves);
results including the next order HBChPT corrections for the spin-flip 
GPs, as calculated in this work, 
and using the leading order predictions for the 
spin-independent GPs (dashed curves);
dispersion relation results~\cite{Pas01,Dre03} (solid curves).
The effect of the anomaly contribution (i.e. $t$-channel $\pi^0$ pole) 
is neglected in all the three model-predictions.
In the lower panels, a comparison is shown 
between the dispersive predictions without the 
anomaly contribution (dashed curves) 
and with the anomaly contribution (solid curves).}
\label{fig:fig3}
\end{figure}

\section{Conclusions}
\label{sec5}

In this work, we calculated the spin-dependent VCS amplitude in 
HBChPT at ${\cal O}(p^4)$. In a low energy 
expansion of the outgoing photon energy, the spin-dependent VCS amplitude 
can be parametrized in terms of 7 generalized spin polarizabilities which 
characterize the response of the nucleon. Nucleon crossing symmetry 
combined with charge conjugation invariance leads to relations among these 
GPs, leaving four independent spin-flip GPs. 
We calculated all seven spin-flip GPs in HBChPT at ${\cal O}(p^4)$ 
and checked the three crossing symmetry relations to this order. 
At order ${\cal O}(p^4)$, no unknown low-energy constants enter the 
theory, allowing us to make absolute predictions for all seven spin-flip GPs 
to this order. Three of the spin-flip GPs at next order have been extracted 
before in Ref.~\cite{KV02} by identifying the term linear in the outgoing 
photon energy in the LEX of the spin-dependent VCS amplitudes. In the present 
work, we calculated within HBChPT at ${\cal O}(p^4)$ 
the quadratic terms in the outgoing photon energy, 
in which the dipole GPs also appear. 
From these quadratic terms, 
we were able to extract the remaining four spin-flip GPs at next 
order in HBChPT. A noteworthy feature of our results is that we provide  
analytical expressions for all spin-flip GPs at next order, analogously 
as has been done at lowest order in Ref.~\cite{HHKD}. 
Another interesting feature arises from 
the crossing symmetry relations which connect the HBChPT amplitudes 
at different orders. As a result, our prediction for the GP $P^{(L1,L1)1}$ 
is formally a N$^3$LO result, and could only be obtained from 
a HBChPT calculation at ${\cal O}(p^5)$, if one were not to use the 
crossing relations. 
\newline
\indent
Our predictions show sizeable corrections to the leading order 
(non-zero) results  
for the spin-flip GPs. The sign of these corrections is such as to move the 
leading order results in the direction of a phenomenological estimate of GPs 
based on dispersion relations. 
As the spin-flip GPs allow for absolute predictions 
at two successive orders in HBChPT without new low-energy constants entering, 
they may also be useful as a testing ground of chiral theories. 
It would therefore 
be desirable to have direct experimental information on the spin-flip 
GPs to test the different predictions. 
\newline
\indent
One combination of the four spin-flip GPs can be extracted 
by doing an unpolarized VCS experiment at a fixed value of $Q^2$ and 
by varying the beam energy. One can extract the other three spin-flip 
GPs through VCS double polarization observables 
with polarized lepton beam and polarized recoil nucleon. We gave predictions 
of these double polarization asymmetries for experimental conditions 
accessible at MAMI. 
We found that the relative effect on these asymmetries due to the 
spin-flip GPs is of the order of 5 \%. Although the measurement of these 
asymmetries requires dedicated experiments, such a measurement would 
challenge the parameter-free chiral predictions at next order 
for the spin-flip GPs presented in this work.

\acknowledgments
This work was supported by the 
U.S. Department of Energy under contract DE-DF0-97ER41048 (C.W.K.),  
and contracts DE-FG02-04ER41302 (M.V.) and DE-AC05-84ER40150 (M.V.).
The work of B.P. was supported by 
the Italian MIUR through the PRIN Theoretical Physics
of the Nucleus and the Many-Body Systems.

\begin{appendix}
\section{Crossing symmetry properties of the VCS amplitudes 
in the covariant basis}
In this appendix, we study the properties of the amplitudes $f_i$ under 
nucleon crossing combined with charge conjugation, 
which transforms $f_{i}(q^2,q\cdot q',q\cdot P)$ into 
$f_{i}(q^2,q\cdot q', -q\cdot P)$. 
First we expand the arguments of the amplitudes $f_i$ as~:
\begin{eqnarray}
f_{i}(q^2,q\cdot q',q\cdot P)&=&\sum_{a,b,c} {\cal C}_{abc}^{i}\cdot
(q^2)^{a}\cdot(q\cdot q')^{b}\cdot(q\cdot P)^{c},\nonumber \\
q^{2}&=&2M_{N}\omega_{0}+2\omega_{0}\cdot \omega'
+(\frac{\omega_{0}}{M_{N}}+1)\cdot\omega'^{2}+{\cal O}(\omega'^{3}),
\nonumber \\
q\cdot q'&=&0+(\omega_{0}-\bar{q}\cos\theta)\cdot\omega'+1\cdot
\omega'^{2}+{\cal O}(\omega'^{3}),\nonumber \\
q\cdot P&=&0+(\frac{-\bar{q}^{2}}{\omega_{0}}
+\bar{q}\cos\theta)\cdot\omega'+1\cdot\omega'^{2}+{\cal O}(\omega'^{3}).
\label{eq:ci}
\end{eqnarray}
Now we obtain~:
\begin{eqnarray}
\mathring{f}_{i}&=&\sum_{a=0}^{\infty} {\cal C}_{a00}^{i}(2M_{N}\omega_{0})^{a},
\nonumber \\
g_{i}&=&\sum_{a=0}^{\infty} {\cal C}_{a10}^{i}(2M_{N}\omega_{0})^{a}\omega_{0}
+\sum_{a=0}^{\infty} {\cal C}_{a01}^{i}(2M_{N}\omega_{0})^{a}(2M_{N}
-\omega_{0})
+\sum_{a=1}^{\infty} {\cal C}_{a00}^{i}\cdot a(2M_{N}\omega_{0})^{a-1}\cdot 2
\omega_{0},\nonumber \\
h_{i}&=&\sum_{a=0}^{\infty}-{\cal C}_{a10}^{i}(2M_{N}\omega_{0})^{a}\bar{q}
+\sum_{a=0}^{\infty}{\cal C}_{a01}^{i}(2M_{N}\omega_{0})^{a}\bar{q} ,
\end{eqnarray}
where ${\cal C}_{abc}^{i}$ are the coefficients which start at $n=0$, 
and ${\cal C}_{abc}^{i}\equiv \sum_{n=0}^{\infty}
\frac{1}{M_{N}^{n}}\cdot {\cal C}_{abc}^{i(n)}$.
\newline
\indent
It was shown in Ref.~\cite{Mainz1} that nucleon crossing symmetry combined 
with charge conjugation leads to the following symmetry properties of the 
amplitudes $f_i$~:
\begin{eqnarray}
f_{i}(q^2, q\cdot q',q\cdot P)&=&f_{i}(q^2,q\cdot q',-q\cdot P),\;
\; i=1,2,5,6,7,9,11,12,  \nonumber \\
f_{i}(q^2,q\cdot q',q\cdot P)&=&-f_{i}(q^2,q\cdot q',-q\cdot P),\;
\; i=3,4,8,10.
\end{eqnarray}
Furthermore, photon crossing leads to the symmetry relations 
at the real photon point~:
\begin{eqnarray}
f_{i}(0,q\cdot q',q\cdot P)= 0 \, , \;\;\; i=7,9.
\end{eqnarray} 
From these symmetry relations, we obtain for the coefficients entering the 
expansion of Eq.~(\ref{eq:ci})~: 
\begin{eqnarray}
{\cal C}_{a01}^{i}&=&0, \;\;\; i=1,2,5,6,11,12, \nonumber \\
{\cal C}^{i}_{a10}&=&{\cal C}_{a00}^{i}=0,\;\;\; i=3,4,8,10, \nonumber \\
{\cal C}^{i}_{a01}&=&{\cal C}_{010}^{i}=0,\;\;\; i=7,9. \nonumber \\
\end{eqnarray}
As a result, we obtain for the LEX of Eq.~(\ref{eq:lexfi})~:
\begin{eqnarray}
g_{i}&=&\sum_{a=0}^{\infty} {\cal C}_{a10}^{i}(2M_{N}\omega_{0})^{a}\omega_{0}
+\sum_{a=1}^{\infty} {\cal C}_{a00}^{i}\cdot a(2M_{N}\omega_{0})^{a-1}\cdot 2\omega_{0},\nonumber \\
h_{i}&=&\sum_{a=0}^{\infty}-{\cal C}_{a10}^{i}(2M_{N}\omega_{0})^{a}\bar{q}
,\;\;\; \mathrm{for} \;\;\; i=1,2,5,6,11,12,
\label{eq:crossing1}
\end{eqnarray}
furthermore 
\begin{eqnarray}
\mathring{f}_{i}&=&0,\,\,
g_{i}=\sum_{a=0}^{\infty} {\cal C}_{a01}^{i}(2M_{N}\omega_{0})^{a}
(2M_N-\omega_{0}),
\nonumber \\
h_{i}&=&\sum_{a=0}^{\infty}{\cal C}_{a01}^{i}(2M_{N}\omega_{0})^{a}\bar{q}
,\;\;\; \mathrm{for} \;\;\; i=3,4,8,10,
\label{eq:crossing2}
\end{eqnarray}
and 
\begin{eqnarray}
g_{i}&=&\sum_{a=1}^{\infty} {\cal C}_{a10}^{i}(2M_{N}\omega_{0})^{a}\omega_{0}
+\sum_{a=1}^{\infty} {\cal C}_{a00}^{i}\cdot a(2M_{N}\omega_{0})^{a-1}\cdot 2\omega_{0},
\nonumber \\
h_{i}&=&\sum_{a=1}^{\infty}-{\cal C}_{a10}^{i}(2M_{N}\omega_{0})^{a}\bar{q},
\; \; \; \mathrm{for} \;\;\; i=7,9. 
\label{eq:crossing3}
\end{eqnarray}
The above results in turn yield relations 
for the coefficients in the heavy baryon 
expansion as in Eqs.~(\ref{eq:hbchpt1},\ref{eq:hbchpt2}).
Since in the heavy baryon expansion, 
$\omega_{0}$ is counted as $n=1$ and $2M_{N}\omega_{0}$ as $n=0$,
we obtain~:
\begin{equation}
g_{1}^{(0)}=g_{2}^{(0)}=g_{5}^{(0)}=g_{6}^{(0)}=g_{7}^{(0)}=g_{9}^{(0)}
=g_{11}^{(0)}=g^{(0)}_{12}=0.
\label{cross2}
\end{equation}
Moreover we know
$$2f_{6}^{(0)}+f_{9}^{(0)}=f_{12}^{(0)}=0,$$
because otherwise the amplitudes $\bar A_{8}$ and $\bar A_{12}$ 
would start at $n=-1$ which is impossible.
Therefore
\begin{eqnarray}
& &2\mathring{f}_{6}^{(0)}+\mathring{f}_{9}^{(0)}=\mathring{f}_{12}^{(0)}=0,
\label{cross3a}
\end{eqnarray}
and analogously
\begin{eqnarray}
2g_{6}^{(0)}+g_{9}^{(0)} &=& g_{12}^{(0)}=0, 
\nonumber \\
2h_{6}^{(0)}+h_{9}^{(0)} &=& h_{12}^{(0)}=0.
\label{cross3b}
\end{eqnarray}
From $h_{12}^{(0)}=0$ it follows that $C_{a10}^{12(n=0)}=0$, and 
from $\mathring{f}_{12}^{(0)}=0$ it follows that $C^{12(n=0)}_{a00}=0.$
Therefore we obtain $g_{12}^{(1)}=0.$
Similarly, one can show that~: 
\begin{eqnarray}
& &2g_{6}^{(1)}+g_{9}^{(1)}=0, \nonumber \\
& &h_{3}^{(0)}=h_{4}^{(0)}=h_{8}^{(0)}=h_{10}^{(0)}=0.
\label{cross3c}
\end{eqnarray}
otherwise $g_{3},g_{4},g_{8},g_{10}$ will start at $n=-1$.
Lastly, Eq.~(\ref{eq:crossing2}) yields the relations~:
\begin{equation}
h_{i}^{(1)}=g_{i}^{(0)}\cdot\frac{\bar{q}}{2M_{N}},
\;\;\; \mathrm{for} \;\;\; i=3,4,8,10.
\label{cross4}
\end{equation}
\end{appendix}

\end{document}